\documentclass{Interspeech}
\usepackage{multirow}
\usepackage{graphicx}
\usepackage{cite}
\usepackage[table,xcdraw]{xcolor}
\usepackage[normalem]{ulem}
\usepackage{pifont}% http://ctan.org/pkg/pifont
\newcommand{\cmark}{\ding{51}}%
\newcommand{\xmark}{\ding{55}}%
\interspeechcameraready

\title{Listen, Analyze, and Adapt to Learn New Attacks: An Exemplar-Free\\ Class Incremental Learning Method for Audio Deepfake Source Tracing}

\author[affiliation={1,2}]{Yang}{Xiao}
\author[affiliation={2}]{Rohan Kumar}{Das}

\affiliation{}{The University of Melbourne}{Australia}
\affiliation{}{Fortemedia Singapore}{Singapore}
\email{yxiao9550@student.unimelb.edu.au, rohankd@fortemedia.com}


\begin{document}

\maketitle

\begin{abstract}
% As deepfake speech becomes common and hard to detect, it is vital to trace its source. Recent work on audio deepfake source tracing (ST) aims to find the origins of synthetic or manipulated speech. However, ST models must adapt to learn new deepfake attacks while retaining knowledge of the previous ones. A major challenge is catastrophic forgetting, where models lose the ability to recognize previously learned attacks. Some continual learning methods help with deepfake detection, but multi-class tasks such as ST introduce additional challenges as the number of classes grows. We propose an analytic class incremental learning method called AnaST. When new attacks appear, the feature extractor remains fixed, and the classifier is updated with a closed-form analytical solution in one epoch. This approach ensures data privacy, optimizes memory usage, and is suitable for online training.
% %This approach protects data privacy, saves memory, and is suitable for online training.
% The experiments carried out in this work show that our method outperforms the current baselines. The related code package will be released soon.
As deepfake speech becomes common and hard to detect, it is vital to trace its source. Recent work on audio deepfake source tracing (ST) aims to find the origins of synthetic or manipulated speech. However, ST models must adapt to learn new deepfake attacks while retaining knowledge of the previous ones. A major challenge is catastrophic forgetting, where models lose the ability to recognize previously learned attacks. Some continual learning methods help with deepfake detection, but multi-class tasks such as ST introduce additional challenges as the number of classes grows. To address this, we propose an analytic class incremental learning method called AnaST. When new attacks appear, the feature extractor remains fixed, and the classifier is updated with a closed-form analytical solution in one epoch. This approach ensures data privacy, optimizes memory usage, and is suitable for online training. The experiments carried out in this work show that our method outperforms the baselines.
\vspace{1mm}
\end{abstract}

\section{Introduction}
In recent years, deepfake generation and detection have drawn much attention~\cite{ASVspoof_journal,VCC2020,zhang2024speaking}. Many global competitions, such as ASVspoof and the ADD challenge series~\cite{asvspoof2019,asvspoof2021,asvspoof5,add2022,add2023}, encourage research on developing advanced and effective deepfake countermeasures. As binary classification (real/fake) has improved over the past few years, studies now focus on tracing the source of audio deepfake. This source tracing (ST) task~\cite{st1,st2,st3,st4,st5} aims to find the system that produced a given speech sample. For example, law enforcement needs to find the source of legal cases, and developers want to protect their intellectual property. Thus, recognizing the audio deepfake algorithm is very important.

Several ST methods have been proposed recently~\cite{st1,st2,st3}, where various model designs and processing techniques are used that work well in controlled test sets. However, speech generation technology has witnessed continuous progress and nowadays deepfake speech sounds very close to real ones.
% has also advanced and now produces fake speech that sounds very real.
This makes it difficult for current ST methods to handle unseen attacks. To address this, researchers have developed two main approaches. The first one uses data augmentation and multi-feature fusion to extract strong audio features, helping the model to manage many types of attacks~\cite{st4,st5,xlsrmamba}. The second one uses continual learning (CL), where the model learns from both previous and new data over time. However, the key problem in CL is {\it catastrophic forgetting}, which makes the model lose previously learned knowledge~\cite{ewc}. Specifically, the ST model must remember previously learned audio deepfakes attacks, while learning unseen new ones.

To overcome catastrophic forgetting, CL~\cite{cl} methods aim to learn new information while retaining the previous knowledge. This approach has been used in speech processing, audio classification, and other fields~\cite{cl3,ucil,cdoa}. For instance, the authors in~\cite{dfwf} propose ``detecting fake without forgetting" (DFWF) for audio deepfake detection. This method uses regularization and a knowledge distillation loss to improve the model. However, audio deepfake detection is a binary classification task, where the number of classes remains consistent while only the input characteristics change. In the ST task, the model must instead learn new classes of unseen attacks over time. This setting is commonly known as {\it class incremental learning} (CIL)~\cite{cil1} which is the most challenging form of CL.

Recently, many CIL methods have been proposed in speech processing. For example, bias correction-based~\cite{bic} and replay-based techniques~\cite{cl2,peng2024dark,chen2024overcoming,AnalyticKWS} store a few samples from past tasks as exemplars to prevent forgetting old knowledge. However, it is often impossible to obtain these samples. For instance, when an industry releases a pre-trained ST model, the public cannot use the private data of that industry for fine-tuning~\cite{addcl1}. Thus, three core challenges arise when updating ST models incrementally. First, the method must handle an increasing number of target classes, not only real/fake assessments. Second, the method must work without using exemplars to protect private data. Third, it must adapt quickly to online training and rapid deployment to handle new attacks. 

In this work, we address the challenges of audio deepfake ST by proposing an exemplar-free CIL method called {\it analytic source tracing}, in short AnaST. Our approach is inspired by analytic learning, which views network training as learning linear layers. Specifically, for the first task, we replace back-propagation with a recursive least-squares (RLS)-like procedure, then freeze the feature extractor and update the classifier using RLS. When new attacks appear, AnaST adapts incrementally by updating the classifier in one epoch with an analytic strategy. This enables the classifier to reach the same output compared to joint training on all tasks. As a result, our model retains previous knowledge while adapting to learn new attacks without using historical examples. Our studies aim to showcase the effectiveness of the proposed AnaST against the existing baselines and also project its efficiency.  
% Experiments show that AnaST improves average accuracy compared to leading baselines and remains efficient in time and parameters.
{\it To the best of our knowledge, this is the first work to apply CIL in ST.} The code for implementation will be released publicly.

\section{AnaST: Analytic Source Tracing}

\subsection{Problem Formulation}

\begin{figure*}[t]
\centering  
\includegraphics[width=0.8\textwidth]{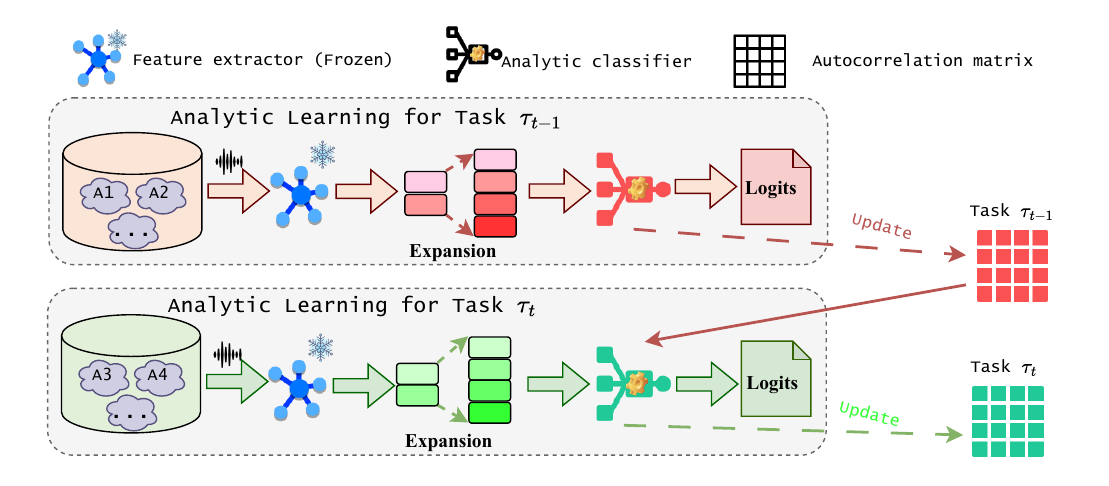}
\vspace{-5mm}
\caption{An overview of the proposed AnaST method for Task \(\tau_t\). We proceed to the class incremental learning stage, where the model adapts by analytic learning for one epoch per new dataset phase, assisted by a correlation matrix (Eq.~\eqref{eq_R_update}) that encodes past knowledge. This process enables the model to learn new tasks while preserving previously acquired information.}
\label{fg1}
\vspace{-5mm}
\end{figure*}

In this work, we examine an ST system that learns different attack categories through a sequence of tasks $\left \{ \tau _{1}, \tau _{2},\dots,\tau _{T} \right \}$. We consider this problem as a CIL scenario, where the system must recognize all attacks from each task, even as new tasks are introduced. For each task $\tau _{t}$, the input data $(x,y)$ follow a distinct distribution $\mathcal{D}_{t}$. Our goal is to train a model $f(x; \mathcal{W} )$ that adapts to new data while preserving its understanding of earlier tasks. Formally, we aim to minimize the cross-entropy loss across all tasks:
\begin{equation}
	\label{eq1}
	\underset{\mathcal{W}}{\mathrm{argmin}} \sum_{t = 0}^{T}\mathbb{E}_{\left ( x,y   \right )\sim \mathcal{D}_{t}  }\left [ \mathcal{L}_{\text{CE}}\left ( y, f(x;\mathcal{W})  \right ) \right ],
\end{equation}
However, storing all past data is impractical due to memory costs and privacy concerns. In addition, simply fine-tuning the model on new data leads to catastrophic forgetting, causing the model to lose knowledge gained from previous tasks.

\subsection{Proposed AnaST Method}

This section describes the AnaST method in detail, along with the initial and analytic learning stages. Our explanation follows the common ST models~\cite{st2,st5}, which include a CNN backbone as an acoustic feature extractor and a linear layer as the classifier. Figure~\ref{fg1} provides an overview of AnaST for task \(\tau_t\).

\subsubsection{Initial training feature extractor}
In the first stage, which is referred as initial training, we use standard back-propagation. During this step, the network is trained on the dataset $\mathcal{D}_{0}$ of task 0 for multiple epochs as the conventional supervised learning. For example, we use common attack types to pre-train the model. After initial training, we obtain a CNN feature extractor with weight \(\mathcal{W}_{\text{cnn}}^{(0)}\) and a linear classifier with weight \(\mathcal{W}_{\text{cls}}^{(0)}\). We aim to adapt this model to handle various attacks in the future. The feature extractor is then frozen to ensure consistency during subsequent stages.

\subsubsection{Analytic learning}

Analytic learning is central to our AnaST approach. We use it from task 0 to task \(\tau_t\).

\noindent {\bf Task 0:}  After obtaining the frozen feature extractor, we change the learning style to analytic learning. In this step, we use the training data $\mathcal{D}_{0}$  from task 0 (with inputs \(x_0\) and labels \(y_0\)) to guide the network toward this style by replacing the standard classifier with an analytic one. The analytic classifier is optimized by least-squares instead of back-propagation~\cite{brmp}.  First, we pass the input through the fixed CNN backbone to get the feature matrix \(\mathcal{F}_0\). Then, we perform a feature expansion by adding an extra linear layer (randomly initialized) to project \(\mathcal{F}_0\) into a higher-dimensional space, resulting in \(\mathcal{F}_0'\). This FE approach is useful for AnaST as it converts the original features into a richer representation without greatly increasing computational costs. The expanded features retain subtle distinctions in speech signals and help in preserving complex patterns. This expanded space, set by a chosen expansion size, provides more information to support analytic learning and improve performance. Finally, we use linear regression to map the expanded features \(\mathcal{F}_0'\) to the label \(y_0\) as:
% The second stage, called analytic re-alignment (Figure~\ref{fg1}(b)), is central to the ACA formulation. In this step, we also use the base training data $D_{0}$ (with inputs \(x_0\) and labels \(y_0\)) to shift the network’s learning toward an analytic-learning style by replacing the classifier with an analytic classifier. First, we pass through the feature extractor CNN (freezed) backbone to obtain the feature matrix \(feat_0\). Next, we perform a feature expansion (FE) process by inserting an extra linear layer to project \(feat_0\) into a higher-dimensional space, resulting in \(feat_0'\). This expanded space, determined by a chosen ``expansion size," provides more information that supports the analytic learning methods and helps the model achieve its best performance. To initialize this FE matrix, we draw each element from a normal distribution, a randomization approach that has shown success in classification tasks. Finally, we use linear regression to map the expanded feature \(feat_0'\) to the label matrix \(y_0\) as:
% \begin{equation}
% \arg\min_{\mathbf{W}_{FCN}^{(0)}} \left\| \mathbf{Y}_0^{train} - \mathbf{X}_0^{(fe)} \mathbf{W}_{FCN}^{(0)} \right\|_F^2 + \gamma \left\| \mathbf{W}_{FCN}^{(0)} \right\|_F^2 
% \end{equation}
\begingroup
\vspace{-2mm}
\begin{equation}
    % \underset\mathbf{W_{mlp}} \mathrm{argmin}
    \underset{{\mathcal{W}}_{\text{cls}}^{(0)}}{\text{argmin}} \left|\left| {y}_0 - \mathcal{F}_0' {\mathcal{W}}_{\text{cls}}^{(0)} \right|\right|^2_\text{FN} + \gamma \left|\left| {\mathcal{W}}_{\text{cls}}^{(0)} \right|\right|_{\text{FN}}^2
    \label{eq_t0}
\end{equation}
\vspace{-3mm}
\endgroup

\noindent where \(||\cdot||_\text{FN}\) indicates the Frobenius Norm, and \(\gamma\) is the regularization of Eq.~\eqref{eq_t0}. The optimal solution to Eq.~\eqref{eq_t0} can be found in the following equation to update the analytic classifier:

\begingroup
\vspace{-2mm}
\begin{equation}
%     \[
% \widehat{{W}}_{\text{FCN}}^{(0)} = \left( {X}_0^{(\text{fe})^\top} {X}_0^{(\text{fe})} + \gamma {I} \right)^{-1} {X}_0^{(\text{fe})^\top} {Y}_0^{\text{train}}
% \]
\widehat{{\mathcal{W}}}_{\text{cls}}^{(0)} = \left( \mathcal{F}_0'{^\top} \mathcal{F}_0' + \gamma {I} \right)^{-1} \mathcal{F}_0'{^\top} {y}_0
\label{eq_t0_so}
\end{equation}
\vspace{-3mm}
\endgroup

\noindent where \(\widehat{{\mathcal{W}}}_{\text{cls}}^{(0)}\) indicates the estimated layer weight of the final analytic classifier layer before outputting the logits. The ST model updates the classifier weights by analytic learning.

\noindent {\bf From task 1 to task \(\tau_t\):} With the network's learning process now aligned to analytic learning (see Eq.~\eqref{eq_t0}), we can advance to CIL using the analytic learning approach. Suppose have access to datasets \(\mathcal{D}_0, \mathcal{D}_1,\dots, \mathcal{D}_{t-1}\). In this case, we can extend the learning task defined in Eq.~\eqref{eq_t0} to incorporate all these datasets, eliminating the risk of forgetting. The linear regression of all expanded features \(\underset{0:t-1}{\mathcal{F}'}\) to the labels $\text{Y}_{0:t-1}$ is:
\begingroup
\vspace{-3mm}
\begin{equation}
% \underset{{W}_{cls}^{(0)}}{\mathrm{argmin}} 
\underset{{\mathcal{W}}_{\text{cls}}^{(t-1)}}{\text{argmin}} \left|\left| 
\text{Y}_{0:t-1}
- 
\underset{0:t-1}{{\mathcal{F}}'}
{\mathcal{W}}_{\text{cls}}^{(t-1)}
\right|\right|_\text{FN}^2 + \gamma \left|\left| {\mathcal{W}}_{\text{cls}}^{(t-1)} \right|\right|_{\text{FN}}^2
\label{eq_t}
\end{equation}
\vspace{-2mm}
\endgroup

\noindent where $\text{Y}_{0:t-1}$ is the block-diagonal matrix whose main diagonal elements are \(y_0, y_1, \dots, y_{t-1}\). And \(\underset{0:t-1}{\mathcal{F}'}\) is formed by stacking the expanded feature matrices. Then the optimal solution to Eq.~\eqref{eq_t} can also be written following the Eq.~\eqref{eq_t0_so} as:
% \begin{equation}
%    \underset{0:t-1}{{y}} =  \begin{bmatrix}
% {y}_0 & 0 & \cdots & 0 \\
% 0 & {y}_1 & \cdots & 0 \\
% \vdots & \vdots & \ddots & \vdots \\
% 0 & 0 & \cdots & {y}_{t-1}
% \end{bmatrix}
% \end{equation}
% and 
% \begin{equation}
%     \underset{0:t-1}{{feat}'} = \begin{bmatrix}
% {feat}_0' \\
% {feat}_1' \\
% \vdots \\
% {feat}_{t-1}'
% \end{bmatrix}
% \end{equation}
% where \(feat'\) is the stack of each task expanded feature. The solution to (5) can be written as:
\begingroup
\vspace{-2mm}
\begin{equation}
\widehat{{\mathcal{W}}}_{\text{cls}}^{(t-1)} = 
\left( 
\sum_{i=0}^{t-1} {\mathcal{F}}_i'{^\top} {\mathcal{F}}_i' + \gamma {I} \right)^{-1} \underset{0:t-1}{{\mathcal{F}}'{^\top}}\text{Y}_{0:t-1}
\label{eq_t_s}
\end{equation}
\vspace{-2mm}
\endgroup

\noindent where \(\widehat{{\mathcal{W}}}_{\text{cls}}^{(t-1)}\) with a column size proportional to task size \(t\). However, in the setting of CIL, we cannot access the previous dataset. So in task \(\tau_t\), we only have the analytic classifier \(\widehat{{\mathcal{W}}}_{\text{cls}}^{(t-1)}\) and dataset \(D_{t}\). We can not directly calculate the \(\widehat{{\mathcal{W}}}_{\text{cls}}^{(t)}\). To solve this, we propose to use the feature autocorrelation matrix (FAuM) to store information about each task’s feature distribution in a compact form. By tracking how features relate to one another, the FAuM allows the system to update its classifier weights using only the new task’s data yet still approximate the result of joint training overall tasks. The FAuM \(\mathcal{R}_{t-1}\) from the task \(\tau_{t-1}\) can be formulate as:

\begingroup
\vspace{-3mm}
\begin{equation}
\mathcal{R}_{t-1} = \left( \sum_{i=0}^{t-1} \mathcal{F}_i'{^\top} \mathcal{F}_i' + \gamma I \right)^{-1}
\end{equation}
\vspace{-2mm}
\endgroup

\noindent The goal of AnaST is to calculate the analytical solution that satisfies Eq.~\eqref{eq_t_s} at task \(\tau_t\) recursively based on \(\widehat{{\mathcal{W}}}_{\text{cls}}^{(t-1)}\), \(\mathcal{F}_{t}'\), and label \(y_t\). When the updated weight \(\widehat{{\mathcal{W}}}_{\text{cls}}^{(t)}\) satisfy Eq.~\eqref{eq_t} with all previous task data, AnaST could reduce forgetting in the sense that the recursive formulation (i.e., incremental learning) gives the same answer with the joint learning. The \(\widehat{{\mathcal{W}}}_{\text{cls}}^{(t)}\) can be calculated by: 

\begingroup
\vspace{-5mm}
            \begin{small}
		\begin{align}\label{eq_w_update}
			 \widehat{\mathcal{W}}_{\text{cls}}^{(t)}
			=\begin{bmatrix}  \widehat{\mathcal{W}}_{\text{cls}}^{(t-1)} -  \mathcal{R}_{t}{\mathcal{F}}_{t}^{\top}\mathcal{F}_{t}' \widehat{{\mathcal{W}}}_{\text{cls}}^{(t-1)}\ \ \   \mathcal{R}_{t}\mathcal{F}_t'{^\top}y_t \end{bmatrix}
		\end{align}
            \end{small}
            \vspace{-5mm}
\endgroup

\noindent where the \(\mathcal{R}_{t}\) could also be recursively calculated by \(\mathcal{R}_{t-1}\) and \(\mathcal{F}_{t}'\) as the equation as:
% \vspace{-1mm}

\begingroup
\vspace{-5mm}
\begin{small}
\begin{equation}
\label{eq_R_update}
	\mathcal{R}_{t} = \mathcal{R}_{t-1} - \mathcal{R}_{t-1} \mathcal{F}_t'{^\top} (I + \mathcal{F}_t'\mathcal{R}_{t-1} \mathcal{F}_t'{^\top})^{-1} \mathcal{F}_t'{^\top} \mathcal{R}_{t-1}
\end{equation}    
\end{small}
\vspace{-5mm}
\endgroup

\noindent For the full proof please refer to the Supplementary Material. Overall, the AnaST approach begins with an initial training phase, where the model learns from the dataset using standard back-propagation. After this training, the classifier is replaced with an analytic classifier, and a fixed feature extractor is used to obtain the expanded feature matrix while saving the feature autocorrelation matrix. Following the analytic learning stage of task 0, the algorithm transitions into a CIL stage. In each CIL task, the model uses the new training data, extracts its feature matrix, updates the FAuM, and finally revises the weight matrix. This process repeats for each incoming task, ensuring that the model adapts to new data while preserving knowledge from previous tasks.
% \noindent For the full proof please see the Supplementary Material. Overall, the AnaST approach begins with an initial training phase, where the model first learns from the \(\mathcal{D}_0\) using conventional back-propagation training. After this training, it replaces the classifier with an analytic classifier and use fixed feature extractor to extract the feature matrix and saves the feature autocorrelation matrix. Following the analytic learning task 0 stage, the algorithm moves into a CIL stage. In each CIL task, it uses the newly received training data for the current task, extracts its feature matrix, updates the FAuM, and finally updates the model’s weight matrix. This process is repeated for each incoming task, ensuring that the model adapts to new data while preserving knowledge from all previously learned tasks.
\vspace{-4mm}
\section{Experiments}
\subsection{Dataset}
We used two datasets in our studies: ASVspoof 2019 LA~\cite{asvspoof2019} and WaveFake~\cite{wavefake}. The ASVspoof 2019 LA dataset has three subsets: training, development, and evaluation. Spoofed utterances are generated by various TTS, VC, and hybrid TTS/VC algorithms. In total, there are 19 attacks, which create 20 classes for the ST task. Because of the uneven attack distribution, we re-split entire ASVspoof 2019 LA into an 80:20 ratio for train and test. The WaveFake dataset is a spoofed audio set generated by seven GAN-based TTS algorithms in English and Japanese. This dataset contains no added noise and includes only two speakers. We also split WaveFake dataset into an 80:20 ratio. 
\subsection{Experimental Setup}

\noindent \textbf{Testbed Model:} We adopt RawNet2 model to evaluate our AnaST method~\cite{rawnet2}. It is an end-to-end neural network that operates on raw waveform inputs. The architecture includes Sinc filters~\cite{sinc}, followed by two Residual Blocks with skip connections on top of a GRU layer to extract representations of the input signal. We use a batch size of 128 and the Adam optimizer with a learning rate of 0.0001. We train for 50 epochs for each task except the AnaST method. For AnaST we only adapt one epoch with an expansion size of 1000. 

\noindent \textbf{CIL Settings: }To cover a range of sequential unseen spoofing attacks, we design two training settings:

For the {\it Single Dataset setting}, we use two datasets and study them separately. In ASVspoof 2019 LA, we pre-train the model using 9 randomly selected attacks and the bonafide class, then simulate incremental learning with 5 tasks, each adding 2 new unique attacks. In WaveFake, we pre-train the model on 3 randomly chosen attacks and simulate incremental learning with 2 tasks, where each task introduces 2 new unique attacks. This setting allows us to evaluate the method on single datasets with different number of tasks.

For the {\it Multi-Dataset setting}, we pre-train the model using data from both ASVspoof 2019 LA and WaveFake data collectively. The pre-training data includes 9 randomly selected attacks and the bonafide class from ASVspoof 2019 LA, along with 1 attack from WaveFake. Then, we simulate incremental learning with 8 tasks, where each task introduces 2 new attacks, alternating between ASVspoof 2019 LA and WaveFake. Thus, this setting tests the method’s performance in a multi-dataset scenario with different sources.

% \begin{itemize}
%     \item Setting 1 (Single Dataset): We pre-train the model on the ASVspoof 2019 LA dataset using 9 randomly selected attacks and the bonafide class. Then, we simulate incremental learning with 5 tasks, where each task adds 2 new unique attacks. This setting is designed to evaluate performance on a single dataset with many tasks. We pre-train the model on the WaveFake dataset with 3 attacks. Next, we simulate incremental learning with 2 tasks, each introducing 2 new unique attacks. This setting tests the method on a single dataset with fewer tasks.
%     \item Setting 3 (Cross-Dataset): We pre-train the model using a mix of data from ASVspoof 2019 LA and WaveFake. The pre-training data includes 9 randomly selected attacks and the bonafide class from ASVspoof 2019 LA, plus 1 attack from WaveFake. For subsequent tasks, each task adds 2 new attacks in alternating order between ASVspoof 2019 LA and WaveFake, for a total of 8 tasks. This setting evaluates the method’s performance across different datasets.
% \end{itemize}

% We use 40-dimensional MFCC as input features. To evaluate our proposed ACA method, we adopt the TC-ResNet-8 model as the backbone. TC-ResNet-8 is a lightweight convolutional neural network developed for KWS tasks on devices with limited computing power. It contains three residual blocks, each composed of 1D temporal convolutional layers, batch normalization layers, and ReLU activation functions. Across these layers, the channel sizes are $ \left \{16,24,32,48  \right \}$, including the first convolutional layer.

\noindent \textbf{Metrics:} We use two metrics to evaluate performance: Average Accuracy (ACC), and Backward Transfer (BWT). ACC refers to the average accuracy across all completed tasks, while BWT measures the impact of learning new tasks on previous ones, indicating whether later learning improves or degrades the earlier knowledge. We also assess efficiency using buffer size that represents the extra memory used to store replay data.

% We also assess efficiency using task training time (TT) and buffer size. TT is the average time required to train each task from all the tasks and buffer size represents the extra memory used to store replay data.

\noindent \textbf{Baselines:} We have built several baselines for comparison.
% Fine-tuning training adapts the model for each new task without using any continual learning strategies and is considered the lower-bound baseline, while joint training uses the entire dataset at once and serves as the upper-bound baseline. We also include two exemplar-free baselines: EWC, a continual learning approach that uses a quadratic penalty to regularize parameters important for past tasks (with importance estimated by the Fisher Information Matrix), and LwF, which updates the model on new tasks while enforcing consistent predictions on existing classes. Additionally, we use two exemplar-based methods: RWalk, which improves EWC by combining Fisher Information Matrix approximation with an online path integral to compute parameter importance and selects historical examples for future training, and DER++, which combines rehearsal and distillation by using both ground truth labels and stored logits to maintain model performance across tasks.
\begin{itemize}
    \item \textbf{Fine-tuning Training:} Adapts the model for each new task without using any continual learning strategy. This method is considered the lower-bound baseline.
    \item \textbf{Joint Training:} Uses the entire dataset at once and serves as the upper-bound baseline.
    \item \textbf{Exemplar-Free CIL Methods:}
    \begin{itemize}
        \item \textbf{EWC~\cite{ewc}:} An approach that uses a quadratic penalty to regularize parameters important for past tasks. The Fisher Information Matrix estimates the importance.
        \item \textbf{LWF~\cite{lwf,dfwf}:} Updates the model on new tasks while enforcing consistent predictions on existing classes.
    \end{itemize}
    \item \textbf{Exemplar-Based CIL Methods:}
    \begin{itemize}
        \item \textbf{Rwalk~\cite{rwalk}:} Improves on EWC by combining Fisher Information Matrix approximation with an online path integral to compute parameter importance while selecting historical examples for future training.
        \item \textbf{DER++~\cite{der}:} Combines rehearsal and distillation by using both ground truth labels and stored logits to maintain model performance across tasks.
    \end{itemize}
\end{itemize}

\begin{table}[ht]
\centering
\caption{Performance comparison in single dataset setting.}
\vspace{-3mm}
\label{tab:combined_table}
\renewcommand{\arraystretch}{1.1}
\setlength{\tabcolsep}{2pt}
\begin{tabular}{>{\raggedright\arraybackslash}m{3.5cm}|c|c|c}
\toprule
\textbf{Method} & \textbf{ACC (\%)} & \textbf{BWT (\%)} & \textbf{BUFFER} \\ 
\midrule
\rowcolor[HTML]{D9EAF2} \multicolumn{4}{c}{\textbf{ASVspoof 2019 LA}} \\ \midrule
Joint (Upper Bound) & 95.37 & - & - \\
Finetune (Lower Bound) & 24.69 & -39.8 & - \\
\hline
EWC & 28.63 & -36.8 & - \\
LWF & 32.11 & -33.7 & - \\
Rwalk & 65.11 & -11.6 & 61M \\
DER++ & 88.06 & -4.6 & 61M \\
AnaST (Proposed) & \textbf{91.68} & \textbf{-3.1} & - \\ 
\midrule
\rowcolor[HTML]{D9EAF2} \multicolumn{4}{c}{\textbf{WaveFake}} \\ \midrule
Joint (Upper Bound)& 99.50 & - & - \\
Finetune (Lower Bound) & 54.80 & -46.8 & - \\
\hline
EWC  & 55.22 & -45.7 & - \\
LWF & 55.98 & -45.3 & - \\
Rwalk & 84.70 & -10.5 & 30M \\
DER++ & 81.10 & -18.9 & 30M \\
AnaST (Proposed) & \textbf{87.85} & \textbf{-10.3} & - \\ 
\bottomrule
\end{tabular}
\end{table}

\vspace{-2mm}
\section{Results and Analysis}
% \usepackage{multirow}
% \usepackage{graphicx}

\subsection{Source Tracing Studies in Single Dataset Setting}

Table~\ref{tab:combined_table} reports the performance of various methods on the two datasets for ST. The joint training results depict the upper bound, indicating high performance on both datasets. In contrast, the fine-tuning method represents the lowest performance with the highest negative BWT indicating forgetting of previously learned information, which stands as the lower bound. 

We now focus on the performance of CIL methods on ASVspoof 2019 LA. The exemplar-free methods (EWC and LWF) achieve low accuracies with severe forgetting, as shown by their negative BWT values. Exemplar-based methods improve performance, with DER++ even achieving 88.06\% ACC. However, they require the use of historical data. Notably, our method, AnaST, achieves 91.68\% accuracy with a BWT of -3.1, closely approaching joint training performance while eliminating the need for exemplars. AnaST only adapts for one epoch, which reduces the reaction time when facing new attacks. 

% Table~\ref{tab:combined_table} shows that AnaST performs the best on the ASVspoof 2019 LA setting compared with other CIL baselines. The joint training reaches 95.37\% accuracy, serving as an upper bound. In contrast, fine-tuning and the exemplar-free methods (EWC and LWF) achieve low accuracies (24.69\% to 32.11\%) with severe forgetting, as shown by their negative BWT values. Exemplar-based methods improve performance, with Rwalk reaching 65.11\% and DER++ achieving 88.06\% accuracy. Notably, our method, AnaST, achieves 91.68\% accuracy with a BWT of -3.1, closely approaching joint training performance while eliminating the need for exemplars.

\begin{figure}[t!]
\centering  
\vspace{-4mm}
\includegraphics[width=\linewidth]{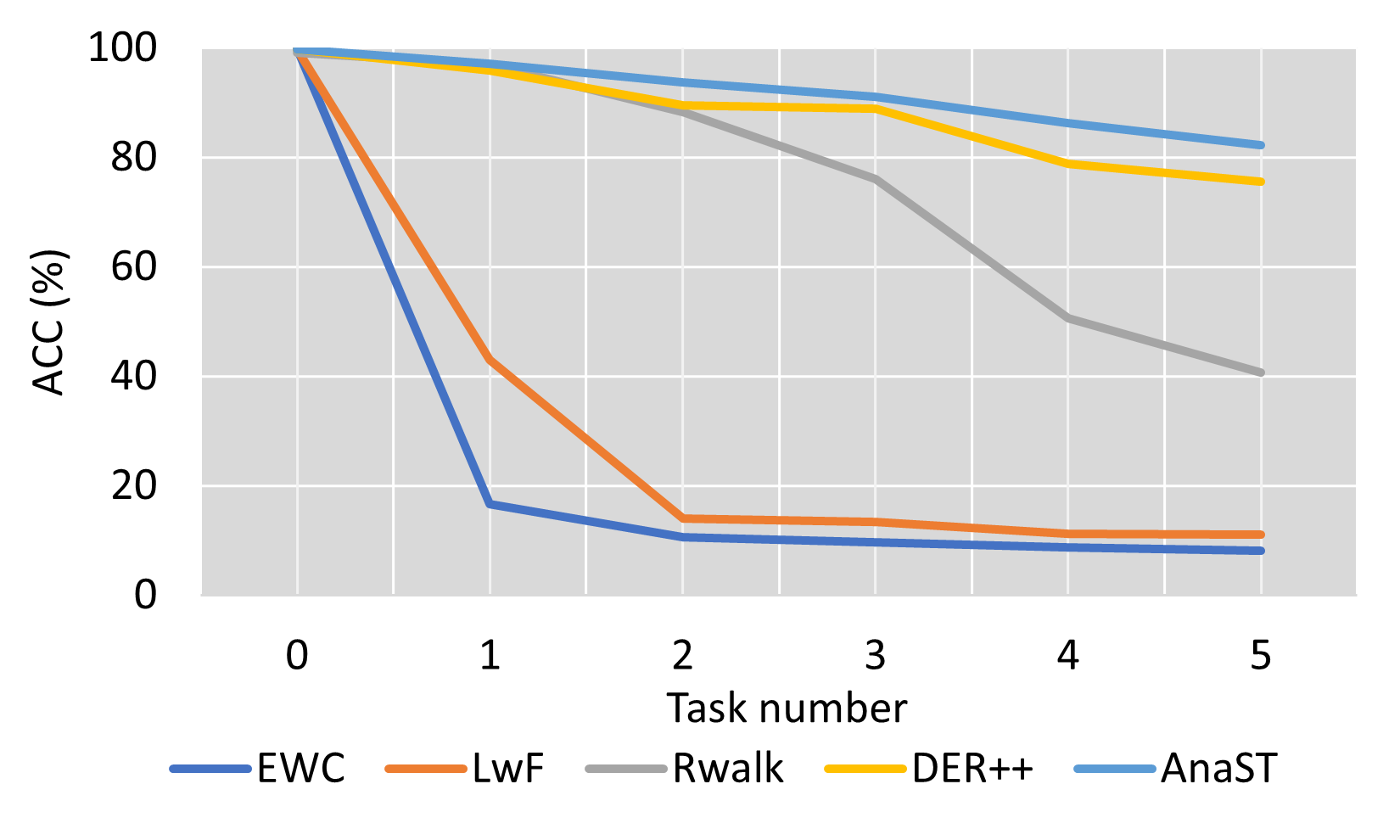}
\vspace{-8mm}
\caption{Task-wise performance in comparison ACC (\%).}
\label{fg2}
\vspace{-6mm}
\end{figure}

For a more detailed analysis, Figure~\ref{fg2} shows the task-wise performance of the CIL methods on ASVspoof 2019 LA, indicating that all methods start with high accuracy on the first task but experience a performance drop as more tasks are added. EWC and LWF experience a swift, deep decline, indicating that they struggle with knowledge retention. Rwalk starts with high accuracy but declines significantly in later tasks. DER++ performs better than other baselines, suggesting that its replay strategy helps to limit forgetting. Finally, the proposed AnaST remains the most stable as it progresses across tasks, suggesting that its analytic structure preserves previously learned information while adding new knowledge.

% For the WaveFake setting as Table~\ref{tab:combined_table} showed, joint training sets a high standard with 99.50\% ACC. Again, fine-tuning and 

On observing the performance of the CIL methods on WaveFake database as reported in Table~\ref{tab:combined_table}, we find that the 
exemplar-free approaches perform poorly (around 55\% ACC) and suffer from severe forgetting. Among the exemplar-based methods, Rwalk obtained better ACC in WaveFake, although still requires a buffer to store past samples. This improved performance is due to RWalk’s regularization, which is well-suited for simple datasets with few classes. Our proposed AnaST method achieves an accuracy of 87.85\% with a BWT of -10.3, demonstrating competitive performance without the need for stored examples. This highlights that AnaST effectively mitigates catastrophic forgetting and adapts well to new attacks while protecting private data.

\subsection{Source Tracing Studies in Multiple Dataset Setting}

% Table~\ref{tab:performance_comparison} reports the performance comparison of various methods under multiple dataset setting. We observe that joint training, as the upper bound, reaches an accuracy of 96.9\%. In contrast, fine-tuning achieves only 20.34\% accuracy with a high negative BWT (-29.9), and the exemplar-free CIL methods EWC and LWF do not improve this performance significantly. However, exemplar-based methods perform better, with Rwalk at 47.01\% accuracy (BWT -12.5) and DER++ at 77.18\% accuracy (BWT -5.5). Notably, our proposed AnaST method obtains 77.85\% accuracy with an improved BWT of -5.4, indicating that it effectively maintains past knowledge while learning new tasks.
Table~\ref{tab:performance_comparison} shows that exemplar-free methods struggle with forgetting, while exemplar-based approaches improve performance by using stored examples. Notably, our proposed AnaST method reaches 77.85\% ACC with a BWT of -5.4, achieving comparable results to the best exemplar-based method without needing any buffer. This indicates that AnaST preserves past knowledge and adapts to new attacks, all while avoiding extra storage overhead. The results also suggest that regularization and our analytic strategy are well-suited for diverse datasets.
% \subsection{Comparable Study for TT}

\begin{table}[t!]
\centering
\caption{Performance comparison in multi-dataset setting.}
\vspace{-3mm}
\label{tab:performance_comparison}
\resizebox{\linewidth}{!}{%
\begin{tabular}{l|c|c|c}
\toprule
\textbf{Method} & \textbf{ACC (\%)} & \textbf{BWT (\%)} & \textbf{BUFFER} \\ 
\midrule
Joint (Upper Bound)     & 96.91  & -        & - \\
Finetune (Lower Bound) & 20.34 & -29.9 & - \\
\midrule
EWC       & 20.63 & -29.6 & - \\
LWF       & 23.62 & -27.2 &  - \\
Rwalk     & 47.01 & -12.5& 61M\\
DER++     & 77.18 & -5.5 &  61M\\
AnaST (Proposed) & \textbf{77.85} & \textbf{-5.4}  & - \\ 
\bottomrule
\end{tabular}}
\vspace{-3mm}
\end{table}

% In terms of training time, our method shows a significant advantage. AnaST requires only one epoch for adaptation and completes training in 666 seconds per task, which is much lower than the 1843 to 2102 seconds observed for the other continual learning methods. Exemplar-based methods such as Rwalk and DER++ also need a buffer of 61M, further increasing their resource demands. These results highlight that while exemplar-based methods improve accuracy, they come with higher computational costs, whereas AnaST achieves competitive performance with lower time and parameter overhead. Moreover, the AnaST method does not require gradient back-propagation for class incremental learning due to analytic learning. This feature makes it ideal for on-device applications.

\begin{table}[t!]
\centering
\caption{Ablation study on the impact of using feature expansion (FE) and regularization \(\gamma\) in proposed AnaST.} 
% The symbol ``\cmark'' indicates the use of FE or regularization, while ``\xmark'' means they are disabled.}
\vspace{-3mm}
\resizebox{0.9\columnwidth}{!}{%
\begin{tabular}{c|c|c}
\toprule
\textbf{Feature Expansion} & \textbf{Regularization ($\gamma$)} & \textbf{ACC (\%)}     \\ \midrule
\xmark                 & \cmark (0.1)            & 75.91\% \\
\cmark           & \cmark (0.1)            & 76.13\% \\
\cmark          & \cmark (0.01)            & \textbf{77.85\%} \\
\cmark           & \cmark (0.001)            & 74.57\% \\
\cmark         & \xmark              & 69.37\% \\ \bottomrule
\end{tabular}%
}
\label{tab:abl}
\vspace{-5mm}
\end{table}

\subsection{Ablation study}

The ablation study in Table~\ref{tab:abl} shows that both feature expansion (FE) and regularization play key roles in our AnaST method. When FE is disabled but regularization is used, the accuracy is 75.91\%. Adding FE with a regularization (0.1) slightly improves accuracy, while setting the regularization to 0.01 with FE gives the best performance of 77.85\%. However, using a very low regularization (0.001) or disabling regularization altogether with FE reduces the accuracy. These results indicate that FE helps create a richer representation and that the proper level of regularization is crucial to maintain optimal performance.

% , highlighting the effectiveness of our method.
\vspace{-2mm}
\section{Conclusion}

In this work, we proposed AnaST, an exemplar-free analytic CIL method for deepfake audio ST. Our method replaces gradient back-propagation with a recursive least-squares procedure and employs feature expansion with proper regularization to update the classifier efficiently while preserving past knowledge. Experiments on the ASVspoof 2019 LA and WaveFake datasets, as well as in a multi-dataset setting, show that proposed AnaST achieves high accuracy with low forgetting and requires significantly fewer computational resources than exemplar-based methods. Ablation studies confirm that both feature expansion and well-tuned regularization are essential for optimal performance. Overall, our work provides a practical and efficient solution for deepfake ST and is well-suited for on-device applications.

\clearpage
\bibliographystyle{IEEEtran}
\bibliography{mybib}

\end{document}